\documentclass[superscriptaddress, nofootinbib, prd]{revtex4}[12pt]
\usepackage{graphicx}% Include figure files
\graphicspath{{figs/}}
\usepackage {amsmath, amssymb, rotating}
\usepackage{hyperref}
\usepackage[usenames,dvipsnames]{color}
\usepackage{mdwlist} %to make compact lists: \begin{itemize*} ... \end{itemize*}
\usepackage{slashed}
\usepackage{verbatim}
\usepackage{mathrsfs}

\def\lsim{\mathrel{\rlap{\lower3pt\hbox{\hskip0pt$\sim$}}
   \raise1pt\hbox{$<$}}}         %less than or approx. symbol
\def\gsim{\mathrel{\rlap{\lower4pt\hbox{\hskip1pt$\sim$}}
   \raise1pt\hbox{$>$}}}         %greater than or approx. symbol

\begin{document}

\title{A Clockwork Axion}

\author{David 
%{\LARGE E}
%$\exists$
%${\cal E}$
%$\mathbb{E}$
%$\mathscr{E}$
E.~Kaplan, {\it p.g.a.}}
\affiliation{Department of Physics \& Astronomy, The Johns Hopkins University, Baltimore, MD  21218}

\author{Riccardo Rattazzi}
\affiliation{Institut de Th\'eorie des Ph\'enom\`enes Physiques, EPFL, Lausanne, Switzerland}

\begin{abstract}
We present a renormalizable theory of scalars in which the low energy effective theory contains a pseudo-Goldstone Boson with a compact field space of $2\pi F$ and an approximate discrete shift symmetry ${\cal Z}_Q$ with $Q\gg 1$, yet the number of fields in the theory goes as $\log{Q}$.  Such a model can serve as a UV completion to models of relaxions and is a new source of exponential scale separation in field theory.  While the model is local in `theory space', it appears to not have a continuum generalization ({\it i.e.}, it cannot be a deconstructed extra dimension).  Our framework  shows that super-Planckian field excursions can be mimicked while sticking to renormalizable four-dimensional quantum field theory.  
We show that  a supersymmetric extension is straightforwardly obtained and we illustrate possible UV completions based on a compact extra-dimension, where all global symmetries arise accidentally as a consequence of gauge invariance and 5D locality.
%Perhaps we should mention the Weak Gravity Conjecture in the abstract.  Nah.
\end{abstract}

\date{\today}
\maketitle

\section{Introduction}

The simplest field theories describing cosmic inflation -- the near-classical slow-rolling of a single field -- which match current cosmological data require a canonically normalized field to take on values larger the the Planck scale \cite{Linde:1983gd}.  In addition, a recent proposal for a solution to the electroweak hierarchy problem -- the relaxion \cite{Graham:2015cka} -- requires field excursions in the early universe much larger than the ultraviolet cutoff of the theory.  While the range of validity of a low-energy effective field theory is applied to the energy, momenta, and masses involved in any physical process, field values are not constrained.  Nevertheless, there is concern that such large field excursions may not be viably generated in a consistent UV completion, especially one encompassing quantum gravity (see for instance \cite{Vafa:2005ui,ArkaniHamed:2006dz}).  Given the implications for inflation and the electroweak hierarchy problem, it is worth looking for existence-proofs of field theories and string theories that produce effectively large field ranges in the infrared.

One way to phrase our question is the following. Consider an ordinary renormalizable quantum field theory in which all scalar fields acquire expectation values $\sim  f< M_{\rm Pl}$ and which contains a pseudo-Goldstone Boson with a compact field space of size $2\pi F$: under what conditions is it  possible to consistently achieve $F\gg f$? For instance how many degrees of freedom, or how much complexity, is needed given a large $F/f$?  We address this question by presenting a simple renormalizable field theory with $N+1$ scalars, where a global symmetry is broken at a scale $f$, and which produces a large field range $F\equiv 3^N f$ for a pseudo-Goldstone Boson in the low-energy theory.
  The model can include couplings to fermions charged under strong gauge groups, and thus produce axion-like potentials with  effective decay constants that range from $f$ to $ 3^N f$.  The interactions in the theory are of nearest-neighbor type and follow a simple pattern, and thus we hope it helps lead to a more complete ultraviolet theory ({\it e.g.}, an embedding in string theory).  We show how to make our construction more plausible by presenting  similar constructions where the global symmetry is an accidental approximate consequence of a higher-dimensional gauge symmetry, and we relate it to multi-axion models already in the literature.  We then show a supersymmetric version can trivially be constructed.

\section{Model}

Our model has $N+1$ complex scalars $\phi_j, j=0,...N$ and the following renormalizable potential:
\begin{equation}
V(\phi) = \sum_{j=0}^N \left( -m^2 \phi_j^\dagger \phi_j + {\lambda\over 4}|\phi_j^\dagger \phi_j|^2\right) + \sum_{j=0}^{N-1} \left( \epsilon \phi_j^\dagger \phi_{j+1}^3 + h.c.\right)
\end{equation}
The first sum of terms respects a $U(1)^{N+1}$ global symmetry, whereas the remainder of the potential breaks this symmetry explicitly to $U(1)$.  Under the unbroken $U(1)$, the fields $\phi_j$, $j=0,1,2,3,...,N$,  have charges $Q = 1, {1\over 3}, {1\over 9},...,{1\over 3^{N}}$.  Note, we ignored the $U(1)^{N+1}$ preserving cross terms, $\phi_j^\dagger\phi_j \phi_k^\dagger\phi_k$, and the first sum preserves a permutation symmetry, but this is only done for simplicity of analysis. Our conclusions will not rely on these assumptions. Note finally that our lagrangian is local in field space, in that it involves only interactions involving neighbours $\phi_j$ and $\phi_{j+1}$, and can thus be associated to a discrete extra dimension. Notice however, as it shall become more clear below, there is no meaningful continuum limit for such an extra dimension.
%
%
% ({\it though we could impose it -- and we have not yet checked what happens otherwise, though I doubt it is important.}).

Taking $\epsilon \ll \lambda < 1$, we can analyze the effective theory at low energies by expanding around an expectation values for the scalars in the limit $\epsilon\rightarrow 0$, namely $\langle |\phi_j|^2 \rangle = f^2 \equiv 2 m^2/\lambda \:\forall\: j$.  Below the scale $\sqrt{\lambda f^2}$, we parameterize the theory in terms of $N+1$ NGBs:
\begin{equation}
\phi_j\rightarrow U_j \equiv f e^{i\pi_j/(\sqrt{2}f)}
\end{equation}
Turning on $\epsilon$, the Lagrangian for the pseudo-NGBs is
\begin{eqnarray}
{\cal L}_{pNGB} &=& {f^2}\sum_{j=0}^N \partial_\mu U_j^\dagger \partial^\mu U_j + \left(\epsilon f^4 \sum_{j=0}^{N-1} U_j^\dagger U_{j+1}^3  +  h.c.\right) + \cdots \\
&=& {1\over 2}\sum_{j=0}^N \partial_\mu \pi_j \partial^\mu \pi_j + \epsilon f^4 \sum_{j=0}^{N-1} e^{i(3\pi_{j+1} - \pi_j)/(\sqrt{2}f)}   +  h.c. + \cdots
\end{eqnarray}
where the ellipsis stands for terms with higher powers of derivatives and/or $\epsilon$.  The terms of order $\epsilon$ generate a potential for the $\pi_j$ fields, except for one linear combination, corresponding to the  Golstone boson of the residual exact $U(1)$.  At the quadratic order, it is given by the $q=3$ case of the general form
\begin{equation}
V^{(2)}=\frac{1}{2}\epsilon f^2 \sum_{j=0}^N (q\pi_{j+1} -\pi_j)^2\,,
\label{massterm}
\end{equation}
corresponding to a mass matrix 
\begin{eqnarray}
M_{ij}^2 = \epsilon f^2 \left(\begin{array}{ccccccccc}1 & -q & 0 & 0 &  &  &  &  &  \\-q & 1+q^2 & -q & 0 &  & . & . & . &  \\0 & -q & 1+q^2 & -q &  &  &  &  &  \\0 & 0 & -q & 1+q^2 &  &  &  &  &  \\ &  &  &  & . &  &  &  &  \\ & . &  &  &  & \phantom{1}. &  &  &  \\ & . &  &  &  &  & \phantom{11}. &  &  \\ & . &  &  &  &  &  & 1+q^2 & -q \\ &  &  &  &  &  &  & -q & q^2\end{array}\right)\,.
\label{eq:massmatrix}
\end{eqnarray}
The eigenmodes are easily found. To get a first idea one can consider  the limit of an infinite matrix, thus ignoring the first and last entries. Equivalently this corresponds to extending the sum in eq.~(\ref{massterm}) from $j=-\infty$ to $j=+\infty$. Our fields in this case can be viewed as living in a discrete infinite extra dimension.
In this case one recovers an exact translational symmetry $\pi_j\to \pi_{j+1}$, and therefore the eigenmodes are simply given by  Fourier modes according to
\begin{equation}
\pi_j=\int_0^{2\pi}\frac{d\theta}{2\pi} e^{ij\theta}\hat \pi(\theta)
\end{equation}
\begin{equation}
V=\frac{1}{2}\epsilon f^2 \int \frac{d\theta}{2\pi} (1+q^2 -2q \cos\theta) |\hat \pi(\theta)|^2
\end{equation}
corresponding to a spectrum filling the band
% to get the eigenvalues of this matrix.the spectrum is easily solved for.  Replacing
%\begin{equation}
%\pi_j = \int_0^{2\pi} d\theta \, e^{i \theta j} \hat{\pi}(\theta) ,
%\end{equation}
%one finds a spectrum
\begin{equation}
m_{\theta}^2 = \epsilon f^2 (1+q^2 - 2q \cos{\theta}), \:\:\: 0 \leq \theta < 2\pi .
\end{equation}
For
 for $q\not = 1$, in particular for $q=3$, the band does not extend to zero mass: the mass gap is $O({\sqrt \epsilon }f)$.  Notice that ${\sqrt \epsilon} f$ represents  the inverse lattice length of our discrete extra-dimension, so that, for $q\not = 1$, 
the Compton wavelength of the modes is comparable to the lattice length and there is no continuum limit. Along with the gapped spectrum, in  the infinite limit there is no normalizable zero mode. A normalizable massless mode however reappears in our  case of finite $N$, which essentially corresponds to a discrete extra dimension of finite length. In that case the mass eigenmodes $\{a_j\}$ satisfying
\begin{equation}
M_{j\ell}^2a_\ell=\lambda a_j
\end{equation}
 consist of a zero mode
 \begin{equation}
 a^{(0)}_j = \frac{\cal N}{q^j}\qquad j=0,\dots,N\qquad\qquad {\cal N}=\frac{1}{\sqrt{ \sum_{j=0}^N 1/q^{2j}}}
 \label{zeromode}
 \end{equation}
 plus massive modes of the form
\begin{equation}
a_j(\theta)= \sin j\theta +A\cos j\theta
\qquad\qquad\lambda=1+q^2-2q \cos\theta\,.
 \end{equation}
The above ansatz for $a_j(\theta)$ and $\lambda(\theta)$ guarantees satisfaction of the eigenvalue equation for $1\leq j\leq N-1$. Satifaction at the boundaries $j=0$ and $j=N$ gives instead the constraints
\begin{eqnarray*}
A&=&-\frac{\sin \theta}{q-\cos \theta}\\
0&=&(1+q^2)\sin(N+1)\theta -q \sin N\theta -q\sin(N+2)\theta\,.
\end{eqnarray*}
The second equation admits $N$ discrete solutions $0<\theta_p<2\pi$ ($p=1,\dots,N$) corresponding to $N$ different eigenvalues $\lambda(\theta_p)$. Leaving their explicit values aside, these eigenvalues are all located in the band we found in the infinite matrix limit: besides the zero mode, all other masses are finite and of order $\sqrt{\epsilon} f$.

Now we see where this is going. In our case, $q=3$, the zero-mode eigenvector is
\begin{eqnarray}
{\vec{a}_{(0)} }^T = {\cal N} \left(\begin{array}{ccccccc}1 & {1\over 3} & {1\over 9} & & ... & & {1\over 3^{N}}\end{array}\right) .
\end{eqnarray}
  This mode has exponentially suppressed overlap, $\sim 3^{-N}$, with whatever operator $\phi_N$ couples to in the UV, and can thus produce an effective decay constant which is exponentially larger than $f$.  For example, if $\phi_N$ couples to fermions, $\psi_N$, charged under a confining gauge group as $y \phi_N \bar{\psi}_{N}\psi_{N}$, the fermions will pick up a mass $y f$, while the zero-mode, $a$, at low energies will pick up the standard axion coupling to the gauge bosons:
\begin{equation}
\sim \frac{a}{32 \pi^2 (3^N f)} H^{\mu\nu}\tilde{H}_{\mu\nu}
\label{eq:UVbroad}
\end{equation}
producing an effective decay constant exponentially larger than $f$ ($H^{\mu\nu}$ is the field strength of the new gauge group).  The classical rolling of this field over large (even super-duper Planckian) field values would correspond to a clockwork of phase-rotations of the $\phi_j$ fields.  With a handful of $\phi_j$ fields, this Goldstone field can be a single large-field inflaton.

To produce a relaxion-type of potential, we will need to add a periodic potential with period $\sim f$.  This can be done by adding a second set of fermions and strong group, and the coupling $y' \phi_0 \bar{\psi}_0 \psi_0$, generating the operator $\sim \frac{\pi}{32 \pi^2 f} G \tilde{G}$.  Below the confinement scales of the two strong groups, the Goldstone's potential becomes:
\begin{equation}
V \sim \Lambda_N^4 \cos{\pi/F} + \Lambda_0^4 \cos{\pi/f}
\end{equation}
where $F \equiv 3^N f$ and $\Lambda_N$ and $\Lambda_0$ are roughly the confinement scales of the two strong groups.  Corresponding to the relaxion potential, the scale $\Lambda_N$ should be at or below the ultraviolet cutoff of the standard model (the scale at which loop contributions to the Higgs mass are cut off), while the second group confines at a scale below the weak scale (and could even be QCD).

\section{Accidental Global Symmetry and the UV}

The structure of the clockwork axion is preserved by an (admittedly bizarre) global $U(1)_{PQ}$ symmetry, and one can ask if such a symmetry should be preserved (by, say, higher-dimensional operators) in light of the expected violation of global symmetries by quantum gravity. Stated otherwise: one could ask if the symmetry at the basis of our mechanism could arise accidentally in a more fundamental theory where all symmetries are gauged. One way to obtain such accidental symmetries is via  locality in extra dimensions.  

One example, using extra-dimensional locality \cite{ArkaniHamed:1998sj}, as applied to the axion \cite{Cheng:2001ys}, is structured as follows. The strong groups live in the bulk of an extra dimension, and our  model's matter, including the scalars and the chiral fermions they couple to,  lives on a brane or hypersurface in an extra dimension.  A chiral partner of the same matter (a copy theory with opposite PQ charges) lives on a different brane (or, for example, on the other orbifold fixed point of an $S^1/{\cal Z}_2$ compactification). Now, because over the entire space, the PQ symmetry is vector-like and thus anomaly-free, the $U(1)$ symmetry can be gauged in the bulk by a $A^{PQ}$ vector.  Anomaly cancelation is insured locally in the 5D theory by including the suitable Chern-Simons terms in the bulk:
\begin{equation}
\frac{C_G}{64\pi^2} A^{PQ}\wedge G^a\wedge G^a+\frac{C_H}{64\pi^2} A^{PQ}\wedge H^\alpha \wedge H^\alpha
\end{equation}
with the correct integers, $C_G$ and $C_H$, to cancel the anomalous representations on each orbifold fixed point, and where $G$ and $H$ represent the field strengths.  While the PQ, $G$, and $H$ 4d vectors are taken to be even under the orbifold, the $A_5$ components are taken to be odd (have Dirichlet boundary conditions), thus projecting out the zero modes.

Taking the compactification scale to be $\pi R \gg 1/M_*$ (where $M_*$ is the 5d cutoff), and assuming the only bulk charged fields are of masses $\sim M_{*}$, the contact terms between the two sectors are suppressed by a Yukawa potential $\sim e^{-\pi R M_*} $, and produce operators such as
\begin{equation}
\frac{{\cal O}^{n}_Q {\tilde{\cal O}}^{\ell}_{-Q}}{M_*^{\ell+n-4}} e^{-\pi R M_*}
\end{equation}
where ${\cal O}$ and ${\tilde{\cal O}}$ are operators from each of the two sectors with dimensions $n$ and $\ell$ and PQ charges $\pm Q$, respectively.  Thus, effectively, there are two symmetries -- one linear combination which is gauged, and the other which is an approximate, accidental global symmetry due to locality.  A spontaneous breaking of the PQ symmetry on one brane causes the (exact) Goldstone to be eaten by the PQ gauge field.  The spontaneous breaking on the second brane produces a near-exact Goldstone (up to exponentially suppressed effects), and thus the global-symmetry is extremely well-preserved. 

%One issue with this UV completion is the fact that a large 5d bulk will dilute the gauge couplings of the strong groups, and thus $\pi R$ cannot be too large if they are to confine at the desired scales.  We set the five-dimensional gauge coupling for a bulk strong group, $g_5$, to its strong coupling value, $g_5^2 \sim 8\pi^2 / M_*$ by naive dimensional analysis (see, for example, \cite{Hebecker:2002vm}), and match it to its 4-dimensional coupling at the compactification scale, $g_4^2 = g_5^2/\pi R\sim 8\pi^2 / (\pi R M_*)$.  Thus, the Yukawa suppression of charged field exchange at the cutoff, $\sim e^{-\pi R M_*}  \sim e^{-8\pi^2/g_4^2}$, is a similar exponential to the contribution of the large (IR) instantons of the strong group, namely 
%\begin{equation}
%\Lambda_{N} = (1/R) e^{8\pi^2 / b g_4^2}
%\end{equation}
%where $b$ is the beta-function coefficient.  Thus, the exponential suppression of the explicit (undesirable) breaking due to the exchange of cutoff-mass charged fields is of the same order {\it in the exponent} as the (desirable) breaking due to IR instantons.  Thus, depending on the UV theory, the undesirable global symmetry breaking can be exponentially suppressed.

Another  partly UV complete theory and structurally analogous to ours   is obtained by extrapolating  the Bi-axion and Tri-axion constructions  in ref. \cite{delaFuente:2014aca} to arbitrary $N$ axions. The set up involves   a 5th dimension again compactified on $S^1/{\cal Z}_2$ of length $\pi R$,  with a $U(1)^{N+1}\equiv U(1)_0\times U(1)_1\dots \times U(1)_N$ gauge theory living in the bulk. The $A_\mu^j$ components ($j=0,\dots,N$) of the gauge fields along the non-compact directions are assumed to satisfy Dirichlet boundary conditions at the boundaries, 
so that no massless extra photon survives compactification. The 5th components $A_5^j$ satisfy von Neumann boundary conditions and give rise to $N+1$ 4D scalars parametrized by the gauge invariant Wilson loops around $S_1$
\begin{equation}
\pi_j \equiv \frac{1}{g_5\sqrt{2\pi R}} \oint dx^5 A_5^j(x^\mu, x^5)\,.
\end{equation}
(Where the normalization ensures $\pi_j$ are canonically normalized when taking a $-F_{MN}F^{MN}/4g_5^2$ 5D gauge kinetic term.).
At the classical level the potential for the $\pi_j$'s vanishes exactly. In the presence of 5D charged matter, a potential will be generated by the Casimir effect, which will depend on the $\pi_j$ via the Aharonov-Bohm phase around the circle.
Given a bulk multiplet with charges $(q_0,\dots,q_N)$ under $U(1)^{N+1}$ the Aharonov-Bohm phase will be
\begin{equation}
\Phi = \frac{1}{f} \sum_{j=0}^N q_j \pi_j\qquad \quad f\equiv \frac{1}{g_5\sqrt{2\pi R} }\,
\end{equation}
and the Casimir energy will be a function of $e^{i\Phi}$. Now, our construction is realized by having $N-1$ bulk fields $\Psi_{j,j+1}$ for $j=0,\dots,N-1$ each with  charges $q_j=1$, $q_{j+1}=-q$ and $q_k=0$ for $k\not = j,j+1$. The quantum numbers of these fields, either boson or fermions, or both, form a moose structure connecting the $U(1)$'s and representing a discrete compact dimension. The effective potential will be a periodic function of the Aharonov-Bohm phases
\begin{equation}
\Phi_{j,j+1} = \frac{1}{f}\left (\pi_j-q\pi_{j+1}\right )
\end{equation}
which are precisely the same combinations we encountered in the 4D model of the previous section. For bulk field masses $m\sim 1/R$ the effective potential will  have  the form
\begin{equation}
V\sim \frac{1}{16\pi^2} \frac{1}{R^4} F\left(\{e^{i\Phi_{j,j+1}}\}\right )
\end{equation}
and around its minimum will give a mass of order $1/(R^2f)$ to $N$ of the scalar, leaving out the same zero mode of  eq.~(\ref{zeromode}). 
 Notice that, depending on the spin and the mass of the link fields, some of the Wilson lines may acquire an expectation value. The details are beyond the scope of this discussion. However it is rather clear that such details in the vacuum dynamics do not affect our main point, which concerns the survival of a massless mode with eigenvector given by eq.~(\ref{zeromode}). Our result for the potential is a direct consequence of the choice of quantum numbers for the light bulk fields. In general one should expect all sort of charges under $U(1)^{N+1}$ corresponding to all possible Aharonov-Bohm phases, and thus no residual massless Goldstone. However, provided the corresponding  fields have a mass $M\gg 1/R$ their contribution to the effective potential will be suppressed by $e^{-2\pi MR}$, with some chance of being negligible.
 
 To make contact with the model of the previous section we can further assume  
the gauge fields of color $SU(3)$ and of some other non-abelian gauge group $H$
propagate in the bulk. We can now consider coupling abelian to non-abelian gauge fields directly via 5D Chern-Simons terms. In particular we can couple $U(1)_0$ to $SU(3)$  and $U(1)_N$ to $H$ by explicitly adding
%\begin{equation}
%\frac{C_0}{64\pi^2}\epsilon^{MNPQR}A^0_MG_{NP}^aG_{QR}^a+\frac{C_N}{64\pi^2}\epsilon^{MNPQR}A^N_MH_{NP}^\alpha H_{QR}^\alpha
%\end{equation}
\begin{equation}
\frac{C_0}{64\pi^2} A^{0}\wedge G^a\wedge G^a+\frac{C_N}{64\pi^2} A^{N}\wedge H^\alpha \wedge H^\alpha
\end{equation}
with $C_0,C_N$ integers.  Notice that in this case, the orbifold boundary conditions are reversed for the $U(1)$s, and their gauge  transformations are  $Z_2$ odd, while $G$ and $H$ are even -- thus, the above CS terms 
%come with $Z_2$ even coefficients and 
are not associated to anomaly inflow at the boundary:   they imply  a coupling to the corresponding Wilson line scalars $\pi_0$ and $\pi_N$
in the low energy effective theory. Focussing just on the surviving axion zero mode $a$ we have
\begin{equation}
\frac{a}{f}\frac{\cal{N}}{64\pi^2}\left ({C_0}G_{\mu\nu}^a\tilde G_{\mu\nu}^a+\frac{C_N}{q^N} H_{\mu\nu}^\alpha \tilde H_{\mu\nu}^\alpha\right )
\end{equation}
so that non-perturbative effects in the two gauge groups give rise to an axion potential modulated over two hierarchically separated periods $f$ and $q^N f$.

One could employ a more elaborate version of this construction to produce an explicit model of the relaxion. For instance, the Higgs boson could be a composite of the confining dynamics of the gauge group factor $H$.  Then all Higgs related parameters would be modulated by the relaxion $a$ but with the slow variation $q^{-N}/f$. We leave such detailed constructions for future work.

One final issue in both UV completions above: one could worry about  the possible relevance of the contribution to the potential of  heavy bulk modes of arbitrary charge. As we said, these effects are controlled by $e^{-2\pi RM}$. It is not realistic to make these effects arbitrarily small by taking $MR$ arbitrarily large. The largest we can hope $M$ to be is the cut-off of the 5D theory $M_*$. In turn $M_*$ is bounded from above by the scale at which the bulk non abelian gauge interactions, $G$ and $H$ become strong. It is an interesting fact\footnote{Which can be explicitly checked using the results in ref. \cite{Hebecker:2002vm}} that one finds
\begin{equation}
2\pi R M_* \sim \frac{16\pi^2}{g^2}
\end{equation}
where $g$ is 4D gauge coupling at the compactification scale $1/R$. Therefore the unwanted corrections to the potential from heavy modes are formally similar to the  non-perturbative 4D effects, such as those associated with confinement.  However given these effects all appear in exponents, there exists a wide range of possibilities. Overall, it does not seem implausible that the IR 4D effects  dominate, but it is not guaranteed. There is however no way to ascertain this in the absence of a full UV completion at the 5D cut-off.

%\section{The Relaxion}

%To fully realize the relaxion, the Higgs must couple to the part of the relaxion potential with periodicity $2\pi F$.  One possibility is that the Higgs is a composite (and perhaps pseudo-Goldstone boson) of  confining fields $\psi_N$ and $ H^{\mu\nu}$.  Of course, a complete such model will require the generation of standard model Yukawa couplings from interactions of the fermion fields in the UV, plus some fermions must carry electroweak quantum numbers (but vector-like, so the reps are anomaly free and the strong group doesn't break electroweak symmetry).  The good news is that because this scale is much higher than the electroweak scale, contributions to electroweak observables and flavor violation are negligible, giving more freedom to build specific models (which is beyond the scope of this paper).

\section{Supersymmetric Clockwork}

One can do a supersymmetric version of the clockwork relatively easily.  Starting with the following superpotential with $3(N+1)$ chiral superfields:
\begin{equation}
W = \sum_{j=0}^N\lambda{S_j (\phi_j \bar{\phi}_j - f^2)} + \epsilon \sum_{j=0}^{N-1} \left(\bar{\phi}_j \phi_{j+1}^2 + \phi_j \bar{\phi}^{2}_{j+1}\right)
\end{equation}
and taking $\epsilon$ small, one can see that the first terms will cause a spontaneous breaking of the approximate $U(1)^{N+1}$ global symmetries (preserving the approximate R-symmetry), and that the $S_j$ fields marry a combination of $\phi_j $ and $\bar{\phi}_j$ fields to produce $2(N+1)$ massive chiral superfields and $N+1$ massless ones.  We parameterize the theory below the mass scale, $\lambda f$, by replacing $\phi_j \rightarrow f e^{\Pi_j/f} , \bar{\phi}_j \rightarrow f e^{-\Pi_j/f} $, where $\Pi_j$ are the (Goldstone) chiral superfields.  The low-energy superpotential becomes:
\begin{equation}
\rightarrow 2 \epsilon f^3 \sum_{j=0}^{N-1} \left(\cosh{(\Pi_j - 2 \Pi_{j+1})/f}\right)
\end{equation}
which produces supersymmetric masses for all but one of the chiral superfields with a mass (not mass-squared) matrix of the form (\ref{eq:massmatrix}) with $q=2$.  Thus, an effective decay constant of $F \equiv 2^N f$ can be produced in the low-energy theory.

\section{Comments}

We have shown that a renormalizable field theory with a modest number of fields can produce an effective field range which is exponentially large compared with the input scales in the theory.  While our construction is ad hoc, it suggests a specific pattern which may be indicative of an even simpler or more complete theory.  
We also suggested UV completions to render all global symmetries accidental, and it would be interesting  to see if similar structures can be made manifest in string constructions and what, if any, the relation is  to existing constructions such as axion monodromy \cite{Silverstein:2008sg,McAllister:2008hb,Kaloper:2008fb,Kaloper:2011jz}. The elements we employ are not totally new and are already present in one form or another in the existing literature \cite{delaFuente:2014aca,ArkaniHamed:2001ca,Kim:2004rp,Bai:2014coa,Choi:2014rja,Higaki:2014pja,Kappl:2014lra,Ben-Dayan:2014zsa}.  However, we have added a twist, and we believe the combination of all those elements and the simplicity of our construction is a novel feature which could open new pathways in model building.

Other than that, we have nothing to say.  Enjoy!

\vskip 1.0 truecm
While this paper was under completion we received a new paper by K. Choi and S. Hui Im \cite{Choi:2015fiu} which presents  different implementations of our same idea.
We would like to thank Prateek Agrawal, Nima Arkani-Hamed, Roberto Contino, David Pirtskhalava, Michele Redi, Prashant Saraswat  and Raman Sundrum for refreshing discussions and encouragement.  DEK acknowledges the support of NSF grant PHY-1214000 and is grateful to the theory groups at EPFL and Stanford University where part of this work was completed.

\end{document}